\documentclass[preprint,showpacs,preprintnumbers,amsmath,amssymb]{revtex4}
\usepackage{graphicx}

\def\physa{Phys. A    }
\def\za{Z. Astrophys.  } 

\begin{document}
\title{Statistics  of cross sections of  Voronoi tessellations }
\author      {M.\ Ferraro}
\affiliation    {Dipartimento  di Fisica Sperimentale
and CNISM,  \\
 via P.\ Giuria 1, I-10125 Turin,Italy }
\email {E-mail:ferraro@ph.unito.it}

\author     {L. Zaninetti}
\homepage   {http://www.ph.unito.it/~zaninett/index.html}
\affiliation    {Dipartimento  di Fisica Generale,
 via P.\ Giuria 1,\\ I-10125 Turin,Italy }
\email{zaninetti@ph.unito.it}
\date {\today}

\begin{abstract}
In this paper we investigate relationships between the volumes of cells of  three-dimensional Voronoi 
tessellations  and the lengths and areas of sections obtained by 
intersecting the tessellation with  a randomly oriented plane.
Here, in order to obtain  analytical results, Voronoi cells are approximated to spheres. 
First,  the probability density function for the lengths 
of the radii of  the sections is derived  
and  it is shown that it  is  related to 
 the Meijer $G$-function; 
its properties are discussed and comparisons are made 
 with the numerical results.
Next the probability density function for the areas of cross sections is computed and compared
 with the results of numerical simulations.

\end{abstract}

\maketitle

\section{Introduction}
Three-dimensional Voronoi tessellations provide a powerful
 method of subdividing 
space in random partitions and have
been used in a variety of fields  such as  computational geometry 
and numerical computing (see for instance 
\cite{Qiang2005}  and the references therein), 
data analysis and compression  \cite{Kanungo2002},
geology \cite{Blower2002}, 
and molecular biology \cite{Poupon2004,Dupuis2011}.
However, in many experimental conditions it is not possible 
to directly observe the cells themselves, just their  planar or linear sections:
thus it is of interest to study the relationships between the geometric properties of
three-dimensional structures and  their lower dimensional sections
\cite{okabe}.

A basic result  has been  proved in \cite{Weygaert1996}: 
the intersection between an arbitrary but fixed plane and a spatial 
Voronoi tessellation is not necessarily a planar Voronoi tessellation. An analysis of some aspects of   the sections of Voronoi diagrams can be found in \cite{okabe}. However, a far as we know, no 
analytical formulas have been derived  for the distributions of  the lengths and areas of the planar sections:  this is precisely the aim of this note.

Following ~\cite{okabe},  a three-dimensional Voronoi diagram will be denoted by 
$\mathcal V$, and section of dimensionality $s$ will be denoted by 
 ${\mathcal V}(s,3)$.

In the next section, some probability density functions used to fit 
empirical, or simulated, distributions of 
Voronoi cells size will be reviewed and in Section
 \ref{secstereology}, results will be presented on the distribution of 
the lengths and areas of  $\mathcal V(2,3)$.

\section{Probability density functions of Voronoi tessellations}
\label{secadopted}
A Voronoi tessellation  is said to be a Poisson Voronoi diagram
(PVD), denoted by  ${\mathcal V}_p$,
if the centers generating the cells  are uniformly distributed. 
In the case of one-dimensional PVD, with average linear density $\lambda$, 
 a rigorous result can be derived, namely that   
the distribution of the lengths of the segments has  
the probability density function

\begin{equation}
\label{eq:ki1}
p(l) = 4\lambda^2 l \exp {(-2\lambda l)},
\end{equation}

\noindent or, by using the standardized  variable $x=l/<l>=\lambda l$ \cite{kiang},

\begin{equation}
\label{eq:nki2}
p(x) = 4x\exp {(-2 x)}.
\end{equation}

No analytical results are known for 
the distribution of the sizes of Voronoi diagrams
in 2D or 3D; 
typically distributions of the surfaces or volumes 
of the Voronoi cells 
are fitted with a $3$-parameter generalized 
gamma PDF of the standardized variable $x$, 
\cite{Hinde1980}, \cite{Tanemura2003},
\begin{equation}
g(x;a,b,c) = 
c \frac {b^{a/c}} {\Gamma (a/c) } x^{a-1} \exp{(-bx^c)} \quad . 
\label{gammag}
\end{equation}

It should be noted that from $g(x;a,b,c)$ other, simpler,  
probability density functions can be derived, that have been also used 
to model empirical  distributions of 
PVD:
for instance, by setting $a=b$ and  $c=1$, 
the  one-parameter distribution proposed in \cite{kiang} results in
 
\begin{equation}
 h (x ;b ) = \frac {b^b} {\Gamma (b)} (x )^{b-1} \exp(-bx),
\label{kiang}
\end{equation}

\noindent with  variance 
\begin{equation}
\sigma_h^2 = \frac{1}{c}.
\end{equation}

Similarly, set 
$a=b=({3d+1})/2$, $c=1$, where $d$ is the dimensionality of the cells:
Eq. (\ref{gammag}) then becomes

\begin{equation}
p(x;d) = \left( \frac{3d+1}{2}\right )^{\frac{3d+1}{2}} x^{\frac {3d-1}{2} } \exp{(-(3d+1)x/2)},
\quad.
\label{rumeni}
\end{equation}

\noindent with variance 
\begin{equation}
\sigma_p^2 = \frac{2}{3d+1}.
\end{equation}

This PDF has been shown to give a good fit for 
Poisson Voronoi diagrams  \cite{Ferenc_2007}, even though it has no free parameters: it will be used in the following to model distributions of the volumes of three-dimensional Voronoi tessellations.

\section{Sections}

\label{secstereology}

\begin{figure*}
\begin{center}
\includegraphics[width=10cm]{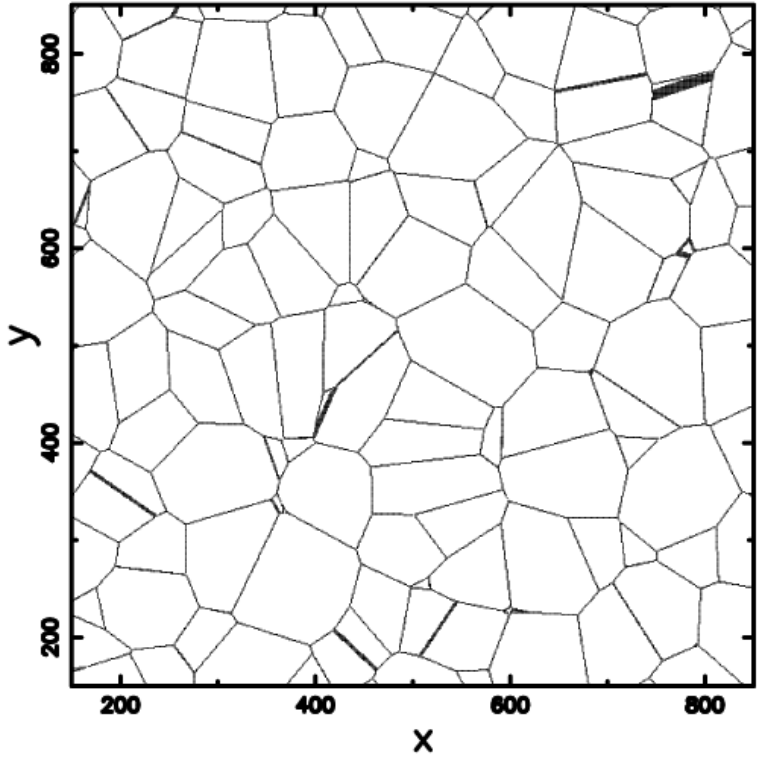}
\end {center}
\caption{
Planar cross  sections of a three-dimensional  Voronoi tessellation.
}
\label{cut_middle}
    \end{figure*}

Consider a three-dimensional Poisson Voronoi diagram  and suppose 
it intersects a randomly oriented plane $\gamma$:
the resulting cross sections are  polygons, as in the  example shown in  
Fig. \ref{cut_middle}. 

In order to obtain analytical results  for the distributions of
the  lengths and areas of ${\mathcal V}_p(2,3)$, some simplifying 
hypothesis is needed: here  the cells will be supposed to be
spheres. This is, admittedly, a rather rough approximation,  but, on the other hand, 
it makes all cuts of the same simple shape, i.e. circles, irrespective of the orientation of the cutting plane. 
As shown in the sequel, this approximation  allows of deriving 
an analytical form  for the distributions of the radii lengths and the areas of cell sections.

Let $f_V(V)$ be  the probability density function 
for the cell volumes, then  
the PDF   $f_R(R)$ for  the lengths of their radii is given by
\begin{equation}
\label{vtor}
f_R\left (R \right )=f_V \left (\frac{4}{3}\pi R^3 \right ) 4 \pi R^2.
\end{equation}  

The use  of Eq. (\ref{vtor}) and  of the standardized PDF (\ref{rumeni}), with $d=3$, leads to  

\begin{equation}
f_R(R) = {\frac {4 \cdot 10^5}{243}}\,{\pi }^{5}{R}^{14}\exp \left (-{\frac{20}{3}}\, \pi \,{R}^{3} \right ).
\label{rumenir}
\end{equation}

\begin{figure*}
\begin{center}
\includegraphics[width=10cm]{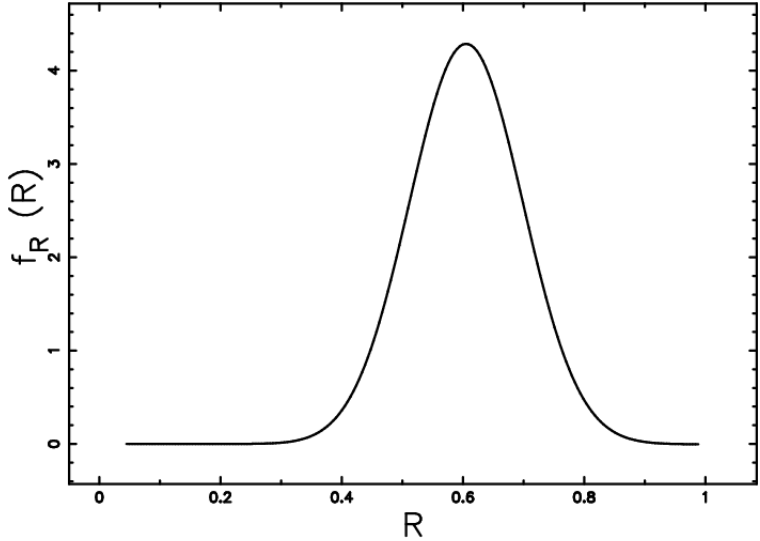}
\end {center}
\caption{
The PDF $f_R$  as a function of $R$.
}
\label{rgreat}
    \end{figure*}

A plot of $f_R$ is shown in Fig.~\ref{rgreat}. 

Consider now the intersection of a sphere of radius $R$ with a randomly oriented plane: the PDF  of  $r$, the length of the  radius of the resulting circle,
is  
\cite{Blower2002} 
\begin{equation}
\pi_c(r)= \int_r^{\infty} f_R (R) \frac{1}{R} \frac{r} {\sqrt{R^2-r^2}} dR.
\label{fund0}
\end{equation}

The probability of a plane's intersecting a sphere is proportional
to $R$ \cite{Blower2002} so, in conclusion, the PDF of $r$ can be written as 
\begin{equation}
f_r(r) = \alpha \int_r^{\infty} f_R(R) \frac{r} {\sqrt{R^2
-r^2}} dR .
\label{fundamental}
\end{equation}

\begin{figure*}
\begin{center}
\includegraphics[width=10cm]{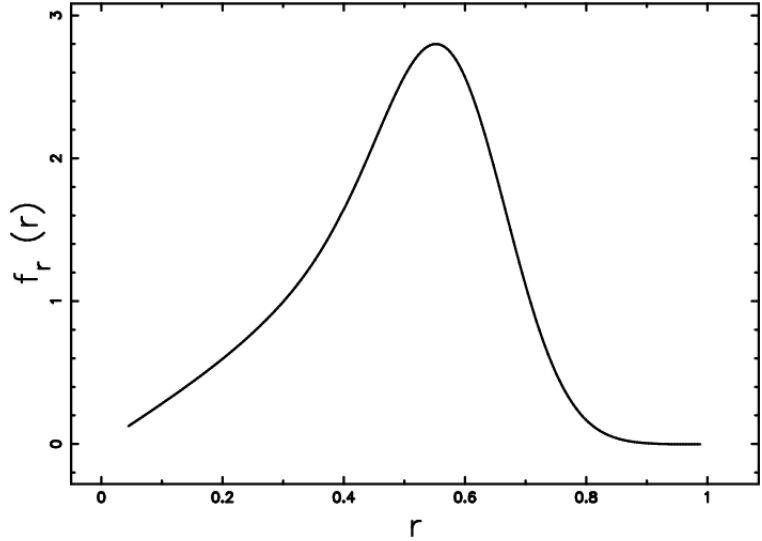}
\end {center}
\caption{
The probability density function $f_r$  as a function of $r$.
}
\label{cut_meijerg}
    \end{figure*}

Insertion of Eq. (\ref{rumenir}) into 
(\ref{fundamental}) results in the formula   
\begin{equation}
\label{intr} 
f_r(r)=\alpha   {\frac {4 \cdot 10^5}{243}}\,{\pi }^{5} \int_r^{\infty}{R}^{14}\exp \left (-{\frac{20}{3}}\, \pi \,{R}^{3} \right )\frac{r} 
{\sqrt{R^2-r^2}} dR.
\end{equation}
The integral (\ref{intr}) can be solved by making use of a computer algebra system such as MAPLE:
the result is 
\begin{equation}
f_r(r)=2/3\,{\alpha}\,\sqrt [6]{3}\sqrt [3]{10}\sqrt [3]{\pi }r
G^{4, 1}_{3, 5}\left({\frac {100}{9}}\,{\pi }^{2}{r}^{6}\,
\Big\vert\,^{5/6, 1/6, 1/2}_{7/3, 2/3, 1/3, 0, {\frac
{17}{6}}}\right),
\label{frmeijer}
\end{equation}
where  $G$ is the Mejier  $G$-function
\cite{Meijer1936,Meijer1941,NIST2010};
the numerical value of $\alpha$, obtained via normalization of $f_r$
is $\alpha=1.649$. A plot of $f_r$ is shown in Fig.~\ref{cut_meijerg}. 
The Meijer  $G$ function is very complicated and Eq. (\ref{frmeijer}) does not lend itself to 
ready interpretation; moreover, at the best of 
our knowledge, it allows no simple approximations 
( see \cite{Luke1969}).

However,  some insight into the  
form of $f_r$ can be gained by considering Taylor expansions. 
In the interval $[0, 0.1]$
$f_r$ is linear, 
$f_r(r) \approx 2.78 r$, while for $r$  close to $1$, 
in the interval $[0.9, 1]$,
it is well approximated  
by a quadratic polynomial: 
\begin{equation}
f_r(r) \approx - 0.006\, \left( r-1 \right) + 0.136
\, \left( r-1 \right) ^{2}.
\end{equation}

The mode occurs in the interval $[0.5, 0.6]$, 
where a close approximation of $f_r$ is given by  
\begin{equation}
f_r(r) \approx 2.48+ 0.58\,r- 95.528\, \left( r- 0.55 \right) ^{
2}- 202.18\, \left( r- 0.55 \right) ^{3}+ 1901.87\, \left( r-
 0.55 \right) ^{4};
\end {equation} 
from this formula the mode $m$ can be 
easily computed: $m=0.553$.

The mean and variance of $f_r$ are 
$\left < r\right >=0.487 $ and 
$\sigma^2_r=0.025$, respectively;
$f_r$ is moderately left skewed (skewness $\gamma_r=-0.522$) 
and is slightly platykurtic 
(excess kurtosis $k_r=-0.111$).
For comparison, note that $f_R$ is more symmetric 
( skewness $\gamma_R=0.014$) and 
has fatter tails (excess kurtosis $k_R=-0.005$). 
 
The  distribution function $F_r$  is
\begin{equation}
\label{dfr}
F_r(r) =
 {\frac {1}{90}}\,{\alpha}\,{3}^{5/6}{10}^{2/3} G^{4, 2}_{4,
6}\left({\frac {100}{9}}\,{\pi }^{2}{r}^{6}\, \Big\vert\,^{1, 7/6,
1/2, 5/6}_{8/3, 1, 2/3, 1/3, {\frac {19}{6}}, 0}\right) {\frac
{1}{\sqrt [3]{\pi }}}.
\end{equation}

The PDF  $f_A$ of the  areas of ${\mathcal V}_p(2,3)$  can be obtained from 
$f_r$  by means of the transformation
\begin{equation}
\label{farea}
f_A(A)=f_r \left ( \left (\frac{A}{\pi}\right )^{1/2}  
\right )\frac{\pi^{-1/2}}{2} {A}^{-1/2},
\end{equation}
\noindent that is, 
\begin{equation}
\label{fareag}
f_A(A)=
 0.549\,\sqrt [6]{3}\sqrt [3]{10}
G^{4, 1}_{3, 5}\left({\frac {100}{9}}\,{\frac {{A}^{3}}{\pi }}\,
\Big\vert\,^{5/6, 1/6, 1/2}_{7/3, 2/3, 1/3, 0, {\frac {17}{6}}}\right)
{\pi }^{-2/3}
\quad . 
\end{equation}

\begin{figure*}
\begin{center}
\includegraphics[width=10cm]{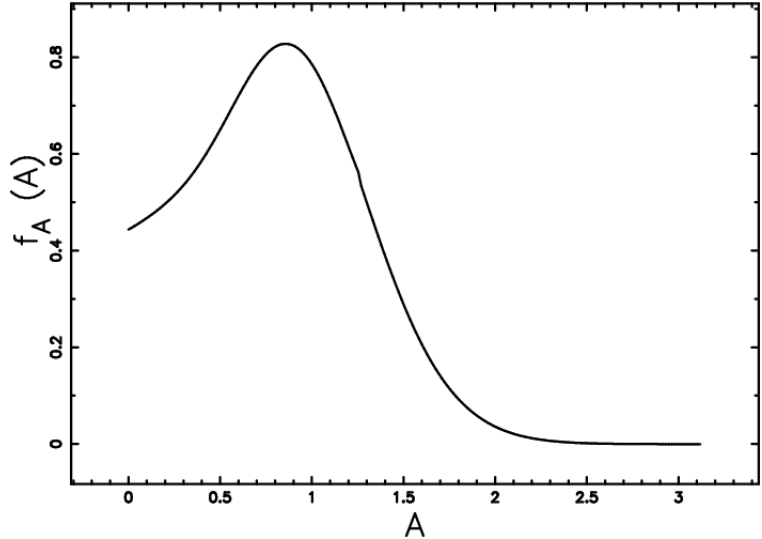}
\end {center}
\caption{
The PDF $f_A$  as a function of $A$.
}
\label{cut_meijerg_area}
    \end{figure*}

Since, for $r$ close to $0$, $f_r(r) \sim r$ from Eq. (\ref{farea}) 
it follows that $f_A(0) \neq 0$ 
and this result has been also verified by numerical simulations we have carried  out 
(see also \cite{okabe} for a further example).
Indeed  it is easy to verify that 
 $f_A(0)=0.443$.
Fig.~\ref{cut_meijerg_area} shows the graph of $f_A$.  

The mode of $f_A$ is $m_A= 0.858 $,
its mean and variance are $\left < A \right >=0.824 $ and  
$\sigma^2_A= 0.204$, respectively; 
$f_A$ is approximately symmetric 
(skewness $\gamma_A=0.278$) and its excess kurtosis is  $k_A=-0.337 $, indicating that 
the transformation from  $f_r$ to $f_A$ yields a PDF with  a lower and broader peak
 and  with shorter and thinner  tails.

The distribution function $F_A$ is given by  
\begin{equation}
\label{dfa}
F_A  =
0.018\,{3}^{5/6}{10}^{2/3}
G^{4, 2}_{4, 6}\left({\frac {100}{9}}\,{\frac {{A}^{3}}{\pi }}\,
\Big\vert\,^{1, 7/6, 1/2, 5/6}_{8/3, 1, 2/3, 1/3, {\frac {19}{6}}, 0}\right)
{\frac {1}{\sqrt [3]{\pi }}}
\quad ;
\end{equation}

A comparison has been carried out  between 
the distribution functions $F_r$, $F_A$, as  
as given by Eq. (\ref{dfr}), (\ref{dfa}) respectively,  
and the results 
of a numerical simulation, performed as follows:  
starting from $300 000$ $3$-dimensional
cells
$100168$ irregular polygons   
were obtained by adding together  results of cuts by  
$41$ triples of  mutually perpendicular  planes. 

The area of the irregular polygons has 
been obtained with the algorithm described in  
\cite{Zaninetti1991}:
briefly, each cut defines a two-dimensional grid and the polygons 
areas we measured by counting the number
 of  points of the grid 
 belonging to a given cell.

\begin{figure*}
\begin{center}
\includegraphics[width=10cm]{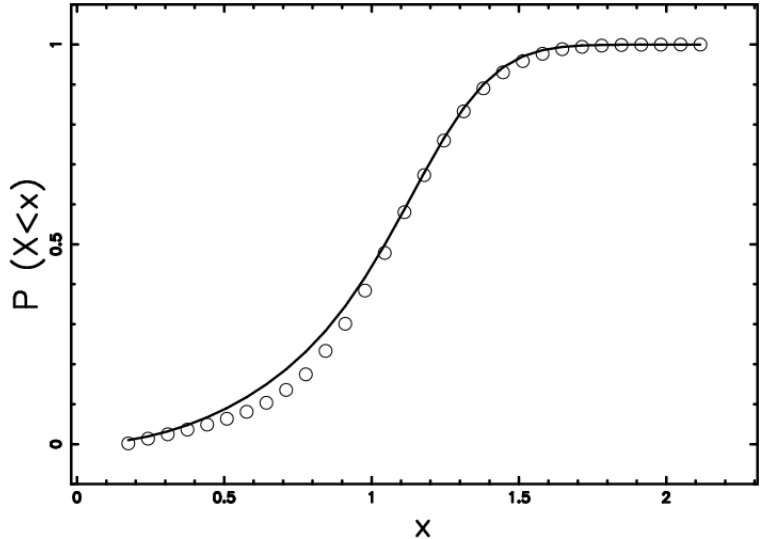}
\end {center}
\caption
{ 
Comparison between theoretical distribution (continuous line)
and data (empty  circles) for the distribution of radii $r$
of 2D cells which result from an intersection with a plane. 
The maximum distance  between the two curves 
 is $d_{max}=0.044$ .
}
\label{frequencies_xyz}
    \end{figure*}

\begin{figure*}
\begin{center}
\includegraphics[width=10cm]{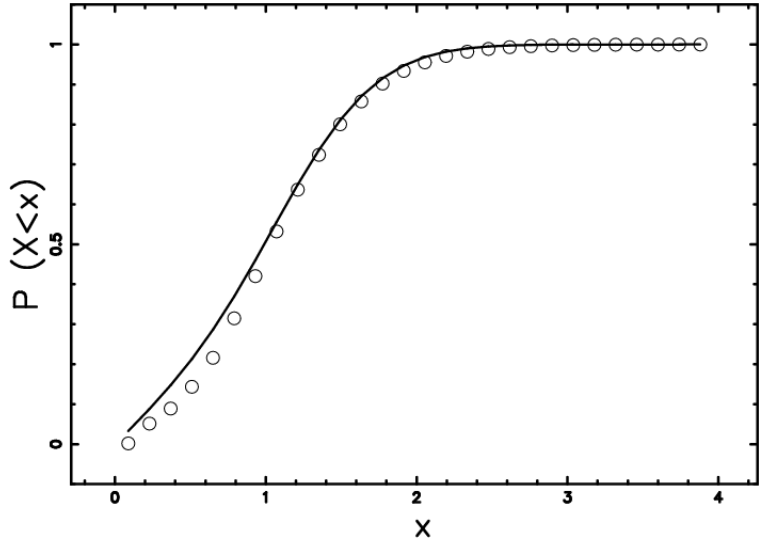}
\end {center}
\caption
{
Comparison between data (empty circles) and theoretical curve
(continuous line) 
of  the distribution of areas.
The maximum distance  between the two curves 
is $d_{max}=0.039$.  
}
\label{area_xyz}
    \end{figure*}

 As concerns the linear dimension, in our approximation the two-dimensional cells were considered 
circles and thus, for consistency, the radius $r$ of an irregular  polygon was defined as 
\begin{equation}
r=\left (\frac{A}{\pi} \right )^{1/2},
\label{empradius}
\end{equation}
that is $r$  is the radius of a circle with the same area of the polygon,
$A$.
The empirical distribution of radii is shown in Fig. \ref {frequencies_xyz} together with
the graph of $F_r$:
the maximum distance between the empirical and analytical curves 
is $d_{max}=0.044$.
Likewise a comparison between $F_A$ and 
the simulation is shown in Fig.~\ref{area_xyz}; 
in this case $d_{max}=0.039$.


\section{Conclusions}

In this paper analytical  formulas have been provided  
which model the distributions of the lengths and areas
 of the planar sections of three-dimensional Poisson Voronoi diagrams:
in particular, it has been  shown that they are related to the Meier $G$ function.

This finding  is consistent with the analytical  results presented in
\cite{Springer1970}, where it is proved that 
nonlinear combinations of gamma variables, such as products or quotients,  
have distributions proportional, or closely related, to the Meijer $G$ distribution.

The analytical distributions $F_r$ and $F_A$ been compared 
with results of numerical simulations:
in evaluating the differences between analytical and empirical distributions  it must be kept in mind  
 that the cells were 
approximated as spheres and that the distributions 
used here have no free parameters that can be  adjusted  
to optimize the fit. 
The results obtained here may be useful for applications in stereology in that
they allow of  predicting the distributions of  linear and planar measures of sections, given 
an arrangement of three-dimensional cells.
This method can also have application to astrophysics, 
namely in the analysis of  the spatial
distributions of the voids between galaxies, that  are experimentally measured  as 
two-dimensional slices  \cite{Vogeley2011} and that can be identified with two-dimensional cuts of a
three-dimensional structure.
For instance, the distribution of the  radius
between galaxies of the Sloan Digital Sky 
Survey Data Release 7
(SDSS DR7) has been reported  in \cite{Vogeley2011}; 
this catalog
contains 1054  radii $r_e$  of voids. 
To compare these experimental data with the results obtained here 
one should know the values of $R_e$, the radii of voids in three dimensions; however 
by making use of standardized variables  
$x=r/{\langle r \rangle}$, 
 $x_e=r_e/{\langle r_e \rangle}$ one obtains
for the standard deviation based  on observations 
$\sigma_{x_e}= 0.26$,  
close to the theoretical value, $\sigma_x=0.32$.
Finally, the $2D$ cuts presented in this paper may be related to 
Laguerre decompositions (weighted Voronoi decomposition)
\cite{Telley1989,Telley1996,Telley1997},
with a weighting factor
depending  on the distance between the between cells 
and the cutting plane.


\end{document}